# Prediction of strong shock structure using the bimodal distribution function

Maxim A. Solovchuk,[1,a] and Tony W. H. Sheu[1,2,b]

[1]Department of Engineering Science and Ocean Engineering, National Taiwan University, No. 1, Sec. 4, Roosevelt Road, Taipei, Taiwan 10617, Republic of China

[2]Center for Quantum Science and Engineering (CQSE), National Taiwan University

A modified Mott-Smith method for predicting the one-dimensional shock wave solution at very high Mach numbers is constructed by developing a system of fluid dynamic equations. The predicted shock solutions in a gas of Maxwell molecules, a hard sphere gas and in argon using the newly proposed formalism are compared with the experimental data, direct-simulation Monte Carlo (DSMC) solution and other solutions computed from some existing theories for Mach numbers M<50. In the limit of an infinitely large Mach number, the predicted shock profiles are also compared with the DSMC solution. The density, temperature and heat flux profiles calculated at different Mach numbers have been shown to have good agreement with the experimental and DSMC solutions

47.40.Nm, 47.45.-n

---

[a] solovchuk@gmail.com

[b] twhsheu@ntu.edu.tw

# INTRODUCTION

A normal shock wave is one example of the highly non-equilibrium flows. The most important parameter that can be used to describe the non-equilibrium properties of the gas is known as the Knudsen number, which can be defined in a shock wave as a relation between the mean free path and the shock thickness. In the shock wave macroscopic properties of the gas can change very rapidly within a short distance, which is about several mean free paths, and the Knudsen number becomes quite large. Strong shock waves post us a computational challenge in the study of stationary highly non-equilibrium flows.

The shock wave structure cannot be described well by fluid dynamic equations in the sense that Navier-Stokes equations [1] give good agreement with the experimental data [2] [3] [4] only at Mach numbers $M < 1.3$. When applying the Burnett and super Burnett equations in the shock prediction some non-physical oscillations were found to appear in the solution even at M=2. [5]

In Grad method [6] and extended irreversible thermodynamics, [7] a large number of equations must be solved to get a reasonable accuracy. [8] Grad's 13-moment method was successful to simulate shock profile below the critical value $M_C = 1.65$. When increasing the number of moments in extended thermodynamics, [7] the solution converges rather slowly. Therefore, a large number of moments is required to get the accurate shock structure at large Knudsen numbers. At Mach numbers $M < 9.36$, for example, one needs up to 15180 equations in extended thermodynamics [7] (506 one-dimensional equations). Until very recently the continuum method for the description of a flow inside the shock wave does not exist. The goal of this study is to develop a system of moment equations for investigating a highly non-equilibrium flow inside the shock wave that is valid in a wide range of Mach numbers.

Good agreement with the experimental measurements was obtained on the basis of bimodal distribution function. [9] However this method fails in the case of low Mach numbers. [10] In order to improve the Mott-Smith method at low and moderate Mach numbers we have proposed the modified Mott-Smith method, which includes a system of fluid dynamic equations.[11] It was shown that we can get the continuous shock structure at all Mach numbers using our theory. Mott-Smith method is able to correctly predict the shock thickness at large Mach numbers. However, the predicted shock wave profiles on the basis of Mott-Smith method disagree with the DSMC simulation results for strong shock waves. [12] It is of interest here to examine whether the proposed system of equations in Ref.[11] is suitable for the description of very strong shock waves and can be applied to improve the Mott-Smith method at an arbitrary Mach number. Recently the problem regarding the structure of very strong shock wave was revisited in connection with the general fluid mechanics development [13] [14] [15]. In this work, the structure of very strong shock waves for the gas of Maxwell molecules, argon and a gas of hard spheres will be studied on the basis of our derived system of equations.

## FLUID DYNAMICS EQUATIONS

The system of fluid dynamic equations to predict the one-dimensional structure of shock wave was derived in Ref.[11] The derived system of equations for the mass density $\rho$, the temperature $T$, the diagonal component of the pressure tensor $P_{XX}$, the vertical component of the heat flow $q_X$ and the new parameter $\overline{q}_X$ has the form given below:

$$\frac{\partial}{\partial t}\rho + \frac{\partial}{\partial x}(\rho U) = 0$$

$$\frac{\partial}{\partial t}U + U\frac{\partial}{\partial x}U + \frac{1}{\rho}\frac{\partial}{\partial x}P_{xx} = 0$$

$$\frac{3k}{2m}\frac{\partial}{\partial t}(\rho T) + \frac{3k}{2m}U\frac{\partial}{\partial x}(\rho T) + (\frac{3k}{2m}\rho T + P_{xx})\frac{\partial}{\partial x}U + \frac{\partial}{\partial x}q_x = 0 \qquad (1)$$

$$\frac{\partial}{\partial t}P_{xx} + U\frac{\partial}{\partial x}P_{xx} + 3P_{xx}\frac{\partial}{\partial x}U + 2\frac{\partial}{\partial x}\overline{q}_x = -\frac{p}{\mu}(P_{xx} - \frac{\rho}{m}kT)$$

$$\frac{\partial}{\partial t}q_x + U\frac{\partial}{\partial x}q_x + 2(q_x + \overline{q}_x)\frac{\partial}{\partial x}U - (\frac{3k}{2m}T + \frac{1}{\rho}P_{xx})\frac{\partial}{\partial x}P_{xx} + \frac{\partial}{\partial x}J_1 = -\frac{2}{3}\frac{p}{\mu}q_x$$

$$\frac{\partial}{\partial t}\overline{q}_x + U\frac{\partial}{\partial x}\overline{q}_x + 4\overline{q}_x\frac{\partial}{\partial x}U - \frac{3}{2\rho}P_{xx}\frac{\partial}{\partial x}P_{xx} + \frac{\partial}{\partial x}J_2 = -\frac{p}{\mu}(\frac{3}{2}\overline{q}_x - \frac{1}{2}q_x)$$

In the above $p = k\rho T/m$ denotes the pressure and $\mu$ is the viscosity. The above system of equations contains two variables given by

$$J_1 = \int d\vec{V}(V_X - U)^2(\vec{V} - \vec{U})^2 f, \quad J_2 = \int d\vec{V}(V_X - U)^4 f \qquad (2)$$

where $f$ is the distribution function of a gas.

To close the system of equations in Eq. (1), we have to prescribe the distribution function. In Ref. [11] we chose the bimodal distribution function.[9] One accounts for the supersonic flow and the other for the subsonic flow:

$$f = f_0 + f_1 \qquad (3)$$

where

$$f_0 = n_0(x)\left(\frac{m}{2\pi kT_0}\right)^{3/2} \exp\left(-\frac{m(\vec{V} - \vec{U}_0)^2}{2kT_0}\right) \qquad (4)$$

Similarly, $f_1$ can be expressed by Eq. (4) by replacing the subscript 0 with the subscript 1. The parameters $T_0, T_1, \vec{U}_1 = (U_1, 0, 0), \vec{U}_0 = (U_0, 0, 0)$ are assumed to be independent of $x$ and $t$. We'll introduce them in the next section through the Rankine-Hugoniot relations.

The expressions of the integrals shown in Eq. (2) are given below

$$J_1 = \frac{n_0(x)}{2}(U-U_0)^4 + 2(U-U_0)^2 n_0(x)V_{T0}^2 + \frac{5}{8}n_0(x)V_{T0}^4$$
$$+ \frac{n_1(x)}{2}(U-U_1)^4 + 2(U-U_1)^2 n_1(x)V_{T1}^2 + \frac{5}{8}n_1(x)V_{T1}^4, \quad (5)$$

$$J_2 = \frac{n_0(x)}{2}(U-U_0)^4 + \frac{3}{2}(U-U_0)^2 n_0(x)V_{T0}^2 + \frac{3}{8}n_0(x)V_{T0}^4$$
$$+ \frac{n_1(x)}{2}(U-U_1)^4 + \frac{3}{2}(U-U_1)^2 n_1(x)V_{T1}^2 + \frac{3}{8}n_1(x)V_{T1}^4$$

where $V_T^2 = 2kT/m$.

## SHOCK STRUCTURE

The shock wave, which is stationary in the steady frame of reference, under current investigation connects the equilibrium states for the density $\rho_0$, velocity $U_0$ and temperature $T_0$ ahead of the shock at $x \to -\infty$ and the equilibrium quantities $\rho_1, U_1, T_1$ behind the shock at $x \to \infty$. It is convenient to employ the dimensionless equations for system (1), where the upstream values are used to define the following dimensionless quantities:

$$\rho' = \frac{\rho}{\rho_0}, U' = \frac{U}{\sqrt{kT_0/m}}, T' = \frac{T}{T_0}, x' = \frac{x}{\lambda_0},$$
$$\pi' = \frac{\pi}{k\rho_0 T_0/m}, q' = \frac{q}{\rho_0(kT_0/m)^{3/2}} \quad (6)$$

In the above, $\pi = p_{xx} - \rho kT/m$ and $\lambda_0$ is the mean free path. The mean free path given in Refs. [1,4,16] will be adopted in this study

$$\lambda_0 = \frac{16}{5\sqrt{2\pi}} \frac{\mu_0}{\rho_0 \sqrt{k/mT_0}} \approx \frac{1}{0.783} \frac{\mu_0}{\rho_0 \sqrt{k/mT_0}} \tag{7}$$

The first three equations, cast in their dimensionless forms (the superscript "prime" in Eq. (6) for the dimensionless variables will be later omitted), in the differential system (1) are as follows:

$$\frac{d}{dx}(\rho U) = 0$$
$$\frac{d}{dx}(\rho U^2 + \rho T + \pi) = 0 \tag{8}$$
$$\frac{d}{dx}(\frac{1}{2}\rho U^3 + \frac{5}{2}\rho T U + \pi U + q) = 0$$

Far ahead of and behind the shock the gas is in equilibrium with $\pi_0 = \pi_1 = 0$ and $q_0 = q_1 = 0$. The dimensionless quantities in front of the shock at $x \to -\infty$ are given by:

$$T_0 = 1, \ \rho_0 = 1, \ U_0 = \sqrt{\frac{5}{3}} M_0 \tag{9}$$

Integration of all the equations in Eq. (8) between the two equilibrium states gives:

$$\rho_1 = \frac{4M_0^2}{M_0^2 + 3}$$
$$U_1 = \sqrt{\frac{5}{3}} \frac{M_0^2 + 3}{4M_0} \tag{10}$$
$$T_1 = \frac{(5M_0^2 - 1)(M_0^2 + 3)}{16M_0^2}$$

It is worth noting that use of the above equations, which are well known as the Rankine-Hugoniot relations, enables us to correctly prescribe the boundary conditions.

The number of equations can be reduced further by integrating the equations in Eq. (8) from the upstream state to an arbitrary location $x$ in the shock. By taking into account Eq. (9), we get:

$$\rho U = \rho_0 U_0$$
$$\rho U^2 + \rho T + \pi = \rho_0 U_0^2 + \rho_0 T_0 \qquad (11)$$
$$\frac{\rho U^3}{2} + \frac{5}{2}\rho TU + \pi U + q = \frac{\rho_0 U_0^3}{2} + \frac{5}{2}\rho_0 T_0 U_0$$

The following relations can be obtained by solving the three equations in Eq. (11):

$$\rho(U) = \sqrt{\frac{5}{3}} \frac{M_0}{U}$$
$$\pi(U,T) = 1 + \frac{5}{3}M_0^2 - \sqrt{\frac{5}{3}}M_0(\frac{T}{U}+U) \qquad (12)$$
$$q(U,T) = \sqrt{\frac{5}{12}}M_0(\frac{5}{3}M_0^2 + 5 + U^2 - 3T) - U(1+\frac{5}{3}M_0^2)$$

Then we substitute the relations in Eq. (12) into the differential system (1) to get the following system of three ordinary differential equations that govern the transport of velocity $U$, temperature $T$ and $\bar{q}$:

$$(3 - \frac{4}{3}\sqrt{15}M_0 U + 5M_0^2)\frac{dU}{dx} + 2\frac{d\bar{q}}{dx} = -\frac{p\lambda_0}{\mu\sqrt{kT_0/m}}(1 + \frac{5}{3}M_0^2 - \sqrt{\frac{5}{3}}M_0(\frac{T}{U}+U))$$

$$\left(\sqrt{\frac{5}{3}}M_0 U^2 + 5\sqrt{\frac{5}{3}}M_0 - \frac{10}{3}UM_0^2 - 2U + 5\frac{\sqrt{15}}{9}M_0^3 - \frac{1}{2}\sqrt{15}M_0 T + 2\bar{q}\right)\frac{dU}{dx} + \frac{d}{dx}J_1 - \frac{1}{2}\sqrt{15}M_0 U \frac{dT}{dx}$$
$$= -\frac{2}{3}\frac{p\lambda_0}{\mu\sqrt{kT_0/m}}\left(\sqrt{\frac{5}{12}}M_0(\frac{5}{3}M_0^2 + 5 + U^2 - 3T) - U(1+\frac{5}{3}M_0^2)\right)$$

$$U\frac{d\bar{q}}{dx} + \frac{d}{dx}J_2 + \left(4\bar{q} + \frac{3}{2}U + \frac{5}{2}M_0^2 U - \frac{1}{2}\sqrt{15}M_0 U^2\right)\frac{dU}{dx}$$
$$= -\frac{p\lambda_0}{\mu\sqrt{kT_0/m}}\left(\frac{3}{2}\bar{q} - \frac{1}{2}\sqrt{\frac{5}{12}}M_0(\frac{5}{3}M_0^2 + 5 + U^2 - 3T) + \frac{1}{2}U(1+\frac{5}{3}M_0^2)\right)$$

The above three equations can be rewritten in the form given below:

$$A \begin{pmatrix} \dfrac{d}{dx}U \\ \dfrac{d}{dx}T \\ \dfrac{d}{dx}\bar{q} \end{pmatrix} = -\dfrac{p\lambda_0}{\mu\sqrt{kT_0/m}} \begin{pmatrix} G_1(U,T,\bar{q}) \\ G_2(U,T,\bar{q}) \\ G_3(U,T,\bar{q}) \end{pmatrix} \quad (13)$$

where $A$ is the $3*3$ matrix with the nonlinear components. The boundary conditions for the investigated system are specified as

$$T_0 = 1, \ U_0 = \sqrt{\dfrac{5}{3}} M_0, \bar{q}_0 = 0 \ \text{at} \ x \to -\infty \quad (14)$$

At $x \to \infty$, we impose

$$U_1 = \sqrt{\dfrac{5}{3}} \dfrac{M_0^2 + 3}{4M_0}, T_1 = \dfrac{(5M_0^2 - 1)(M_0^2 + 3)}{16M_0^2}, \bar{q}_1 = 0 \quad (15)$$

After solving Eq. (13) to get the explicit expressions of three derivatives, we can then solve the coupled first-order ordinary differential equations to get the solutions that connect the information at two fixed ends (boundary conditions at $x \to -\infty$ and $x \to \infty$). The system of equations was derived on the basis of the Boltzmann collision integral for the Maxwell molecules.[17] The corresponding viscosity, which is proportional to the temperature, follows the expression given below with $s = 1$:

$$\mu = \mu_0 \left(\dfrac{T}{T_0}\right)^s. \quad (16)$$

For other interaction potentials the viscosity takes the same form just with an adjustment of the exponent $s$.[13,18,19] For example, $s = 1/2$ is chosen for the hard sphere and $s \approx 0.72$ for the argon.[1,13,18] The hard sphere and the Maxwellian gases are the theoretical gases which can be viewed as the limiting cases of a real gas, since for almost all real gases $0.5 < s < 1$. According to Eqs. (7) and (16), one gets

$$\frac{p\lambda_0}{\mu\sqrt{kT_0/m}} = \frac{\rho T^{1-S}}{0.783}. \tag{17}$$

## COMPARISON STUDY AND DISCUSSION OF RESULTS

To compute the solutions of temperature and velocity in shock profiles from the proposed system of ordinary differential equations in Eq. (13), subject to the boundary conditions (14) and (15), the computational domain is descretized by $N+2$ positions at $x_i$ with $i = 0,1,2...,N+1$. The following approximation under the constant step size $\Delta x$ is used at the nodal point $i$:

$$\left.\frac{dT}{dx}\right|_i = \frac{T_{i+1} - T_{i-1}}{2\Delta x}$$

Calculation of the solutions at positions $x_1$ and $x_N$ requires knowing the field values at $x_0$ and $x_{N+1}$, which are given by (14), (15). One needs therefore to derive 3N coupled algebraic equations for the N unknown values of $U$, $T$ and $\bar{q}$. The resulting nonlinear system was solved with the appropriate $\tanh(x)$ curve being considered as an initial guess for the velocity and temperature (similar to Ref. [11,20]). The predicted temperature and density are presented in the normalized forms $\frac{T-T_0}{T_1-T_0}$ and $\frac{n-n_0}{n_1-n_0}$. One of the main parameters which can well describe the shock profile is the shock thickness, which is defined as

$$\delta = \frac{\rho_1 - \rho_0}{\max(\frac{\partial \rho}{\partial x})}$$

The inverse thickness can be derived from Eq. (18) as $\frac{\lambda}{\delta} = \frac{\alpha}{4}$ according to the Mott-Smith theory. Another quantity is the temperature-density separation $\Delta_{T\rho}$, which is the distance between the two points at which $T = 0.5$ and $\rho = 0.5$, respectively.

**Shock wave results for the Maxwell gas**

In Fig. 1 we compare our results of the temperature profile with the results of DSMC simulation [21] for Maxwell molecules, Navier-Stokes results, and Mott-Smith results at $M = 35$. It is important to point out here that the temperature profile in Fig. 1 shows its maximum within the shock layer, which can't be predicted by Mott-Smith theory and Navier-Stokes equations. The temperature profile becomes non-monotonic at a Mach number $M > 3$. It is well known that such a temperature profile is not the result of a mathematical artifact but is rather the consequence of the atomistic dynamics.[22,23,24] It is worth noting that the predicted temperature-density separation by Mott-Smith theory is smaller in comparison with the DSMC value. The temperature-density separation predicted by Mott-Smith theory[9] is $\Delta_{T\rho} = 20.1\lambda_0$ at $M = 35$, while in our theory we get $\Delta_{T\rho} = 25.7\lambda_0$, which agrees with the DSMC value.[21] For the Navier-Stokes equations the value of $\Delta_{T\rho}$ is $12.5\lambda_0$.

In Fig. 2 the predicted values of the temperature-density separation $\Delta_{T\rho}$ are compared with the Monte-Carlo simulation results.[4,21,25] Mott-Smith theory only gives good agreement with the DSMC simulation in the range of Mach numbers $2.2 < M < 2.7$. The Burnett results of Fiscko and Chapman [21] gives a better agreement with the DSMC simulation result than the Navier-Stokes result. Note that to get a stable solution one term has been deleted from the viscous stress tensor in the Burnett equations. Our results agree well with the DSMC calculation in the entire range of $1 < M < 50$.

**Shock wave results for the argon**

Next, shock parameters are compared with the Monte-Carlo simulation results for argon. Figs. 3, 4 show the temperature and density profiles calculated from the DSMC simulation,[21] Navier-Stokes, Burnett[21] and our proposed equations for the argon investigated at the Mach number $M = 35$. The normalized density of our solutions at the coordinate origin $x = 0$ is exactly 0.5 at any Mach number. Argon is modeled with the value $s = 0.72$ for the viscosity exponent in the above mentioned constitutive equation for $\mu$. Our results are all the time in excellent agreement with the DSMC simulation results for both of the density and temperature profiles. In Fig. 5 the predicted values of the temperature-density separation are compared with the Monte-Carlo simulation results. Employment of Burnett theory [21] gives good agreement with the DSMC simulation results only in the range of small Mach numbers. Our results have, however, good agreement with the DSMC simulation results in the whole range of the Mach numbers $1 < M < 50$.

**Shock wave results for the hard sphere**

Before discussing the results for the hard spheres, we consider also the limit of our equations. The derivation of our proposed fluid dynamic equations is based on the Boltzmann collision integral for the Maxwell molecules. The extension to a more general particle interaction case via the viscosity exponent in Eq. (16) involves only a first approximation. The full system of equations for the hard sphere must include the hard sphere collision integrals of higher moments.

Figs. 6, 7, 8 show the density, temperature and heat flux profiles calculated from the NEMD simulation, [23,26] Navier-Stokes, Holian-Mareschal, and our proposed equations for the hard sphere gas investigated at $M = 134$. In Fig. 6 we have added the density profile, calculated from the Holian and Mareschal equations, [14] that was not presented in Ref. [26]. Quite recently Holian and Mareschal have modified Navier-Stokes equations. [14] One equation for the heat flux vector was derived for the case of a very strong shock wave. They also introduced two free parameters $\delta_1$ and $\delta_2$ that are connected with the Burnett nonlinearity in the conductivity and temperature relaxation. Holian-Mareschal results obtained with the inclusion of nonlinear Burnett conductivity as well as the temperature relaxation agree well with the NEMD results in the upstream part. Their predicted results differ, however, from the NEMD results in the downstream region. In our predicted results one can see also the good agreement with the NEMD simulation results for temperature and heat flux profiles in the downstream shock region. In the upstream region there is only a small disagreement with the NEMD results. We can, as a result, conclude that both of Holian-Mareschal and our results agree with the NEMD simulation result. Our predicted density profile agrees excellently with the NEMD simulation result in both downstream and upstream regions. Holian and Mareschal presented their model only for the case of strong shock waves, while our model can be applied to the whole range of the Mach numbers. Recall that the continuum modeling of a hard sphere gas via the viscosity exponent in Eq. (16) can be considered only as the first approximation. A more intensive investigation regarding the Boltzmann collision integral

for the hard sphere gas is needed to get a better approximation. For the case of Maxwell molecules and a real gas, which is now chosen to be the argon, our predicted results are in excellent agreement with the DSMC data. In Fig. 9 we compare our results of the inverse density thicknesses for the Maxwell molecules and a hard sphere gas with the Monte-Carlo simulations results, [25] [27] Navier-Stokes and Burnett [27] results. Our results are seen to have good agreement with the Monte-Carlo simulations results at all investigated Mach numbers.

**Discussion of results**

We have made a modification on the Mott-Smith method and have derived the system of fluid dynamic equations for the flow inside the shock wave. The Mott-Smith solution is qualitatively correct for $2 < M < 3$. [11] At other higher Mach numbers their predicted errors can be quite large. On the contrary, our presented theory can predict the solution that is in good agreement with the DSMC, NEMD simulations and experimental results for a gas of Maxwell molecules, hard sphere gas, and argon in the range of $1 < M < 50$. In contrast to the solutions predicted by Navier-Stokes theory and other fluid dynamic equations mentioned earlier, the results of our model are in good agreement with the DSMC and experimental results in a much wider range of Mach numbers. As far as we know there is no such a fluid dynamic model that can provide the solutions comparable to the Monte-Carlo simulation results and experiments in the entire range of Mach numbers $M < 50$.

In the derivation of governing equations we used the Mott-Smith distribution function to close the differential system. According to the recent molecular dynamic [28] and direct

Monte-Carlo simulations, [18] as well as the experimental work [29] the main conclusions [30] about a bimodal structure of distribution function in a shock region are correct. In the upstream and downstream region of the shock wave the bimodal distribution function can describe the exact solution of the problem. Inside the shock wave use of the bimodal function gives only the approximate solution. We are interested only in the macroscopic properties inside the shock wave; therefore small errors in the distribution function are not so significant.

Mott-Smith considered only one moment equation to determine the density inside the shock wave. Other macroscopic properties (temperature, heat flux and pressure) are calculated from the appropriate moments of the bimodal distribution function. In our study the macroscopic variables are calculated directly from the system of fluid dynamic equations. By adding two additional moment equations to the Mott-Smith method, we can get some additional insights into the behaviors of the temperature, heat flux and pressure in the whole range of Mach numbers.

## CONCLUSION

We have derived a system of fluid dynamic equations on the basis of the Mott-Smith method for exploring the structure of very strong shock waves. Our predicted temperature, density and heat flux profiles are seen to agree well with the experimental data and the DSMC simulation results in the entire investigated Mach numbers range $M < 50$ for the three investigated gases: the Maxwell gas, a gas of hard spheres and the argon. In the limit of an infinitely large Mach number, the predicted shock profiles are in good agreement with the NEMD solution for a gas of hard spheres. Our system can be considered as an extension of the Navier-Stokes equations, which are valid only at small Mach numbers. In order to get a solution with better agreement with the experimental result many moments are required in extended thermodynamics. With the Mott-Smith closure, a fairly good agreement with the experimental and the Monte-Carlo simulation results can be obtained even from a differential system with much fewer equations. Mott-Smith method was applied to different shock formation problems, including the shock structure in dense gases,[11,15] gas mixtures,[31,32] relativistic shocks,[33,34] and plasma.[16] The proposed model can be also applied to simulate the problems involving polyatomic gases, gas mixtures, plasma and problem in astrophysics.

## ACKNOWLEDGEMENT

This work was supported by a research grant from the National Science Council of Republic of China with the contract number NSC-97-2221-E-002-250-MY3.

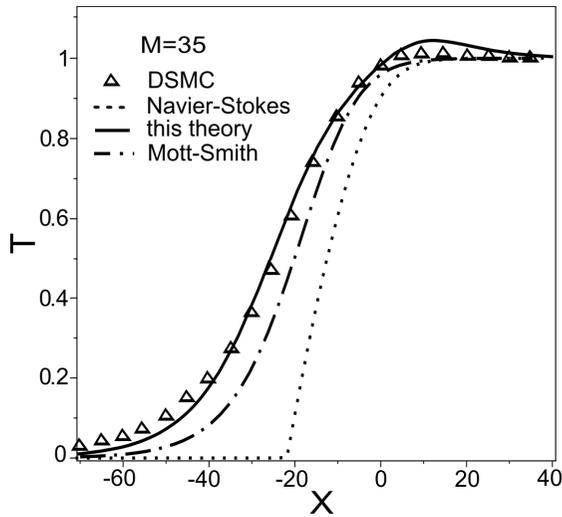

Fig. 1. Temperature profiles plotted as the function of distance. The currently predicted results are compared with those based on the theories of Navier-Stokes, Mott-Smith and DSMC.

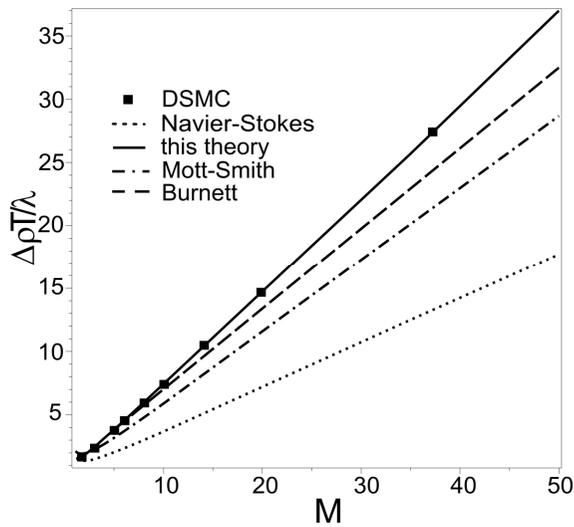

Fig. 2. Comparison of the predicted values of the temperature-density separation, which are plotted against the Mach number. Squares – DSMC results of the Pham-Van-Diep [4], Nanbu [25] and Fiscko [21]

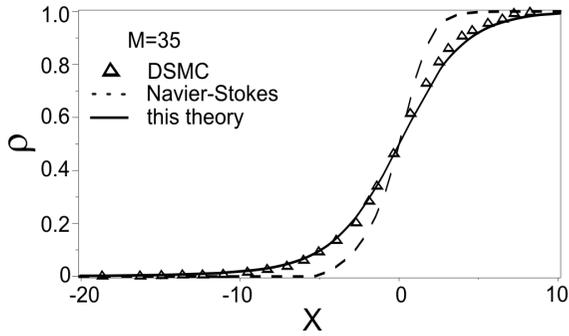

**Fig. 3. Density profile plotted as the function of distance. Comparison of the currently predicted density profile with the DSMC and Navier-Stokes simulation results against *x* at *M=35*; argon, s=0.72.**

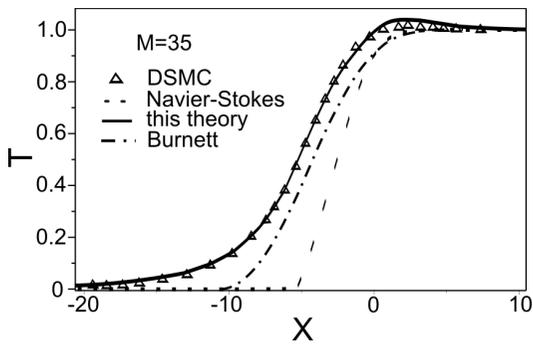

**FIG. 4. Temperature profile plotted as the function of distance. Comparison of the currently predicted temperature profile with the DSMC , Burnett and Navier-Stokes simulation results against x at M=35; argon, s=0.72**

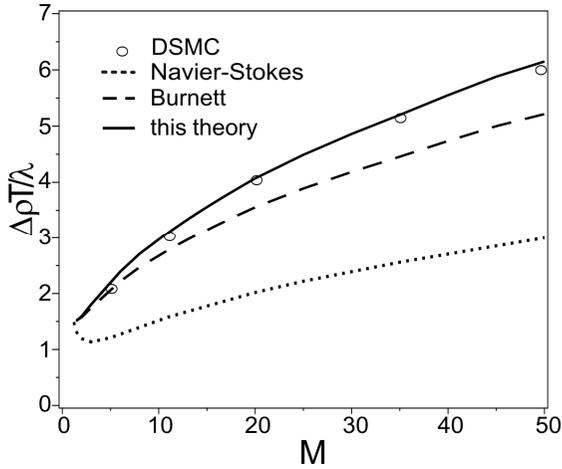

**FIG. 5. Comparison of the predicted values of the temperature-density separation, which are plotted against the Mach number; argon, s=0.72.**

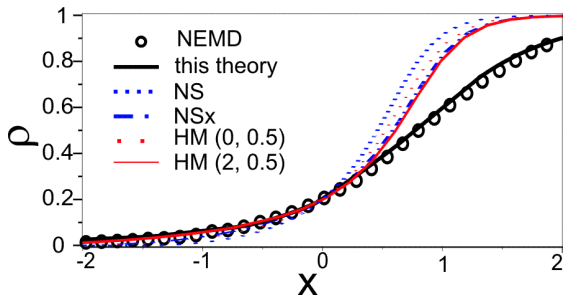

**Fig. 6. (Color online) Density profile plotted as the function of distance. Comparison of the currently predicted temperature profile with the non-equilibrium molecular-dynamics, [23] [26] NSx=NS with the T-dependence of the transport coefficients being replaced by $T_{xx}$, HM (0,0.5) is Holian-Mareschal result with the temperature relaxation only, HM (2,0.5) includes the nonlinear Burnett conductivity as well as the relaxation, Navier-Stokes simulation results against x at M=134; a hard sphere gas, s=0.5.**

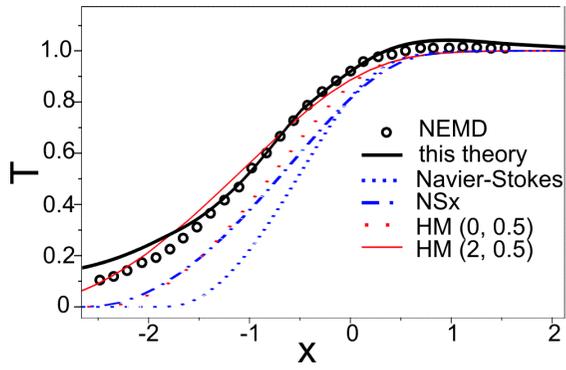

**FIG. 7. (Color online) Comparison of the predicted temperature profiles plotted as the function of distance. M=134; a hard sphere gas, s=0.5. Notation - see Fig. 5.**

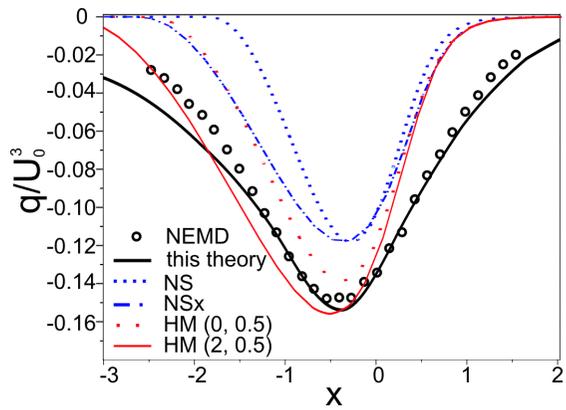

**FIG. 8. (Color online) The predicted heat flux profile plotted as the function of distance. M=134; a hard sphere gas, s=0.5. Notation - see Fig. 5.**

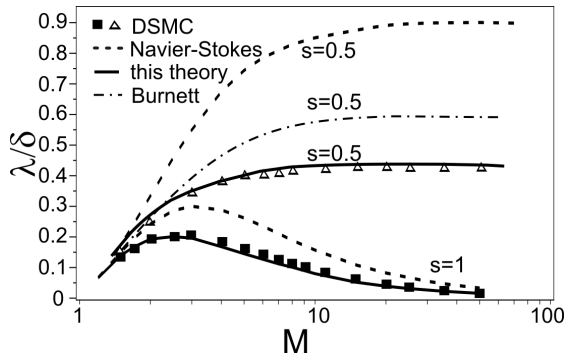

**FIG. 9 Comparison of the computed inverse density thicknesses, which are plotted against the Mach number, for two monatomic gases, Maxwell molecules and a hard sphere gas . DSMC [25] [27]**